\documentclass[11pt,showpacs,preprintnumbers,superscriptaddress,amsmath,amssymb,nofootinbib]{revtex4}
\usepackage{graphicx}
\usepackage{dcolumn}
\usepackage{bm}
\usepackage{amssymb}
\usepackage{amsmath}
\usepackage{epsfig}    
\usepackage{color}
\usepackage{slashed}
\usepackage{hhline}
\usepackage{ulem}

\def\be{\begin{equation}}
\def\ee{\end{equation}}
\newcommand{\bea}{\begin{eqnarray}}
\newcommand{\eea}{\end{eqnarray}}
\newcommand{\nn}{\nonumber}

\begin{document}


\title{Type-II seesaw of a non-holomorphic modular $A_4$ symmetry}


\author{Takaaki Nomura}
\email{nomura@scu.edu.cn}
\affiliation{College of Physics, Sichuan University, Chengdu 610065, China}

\author{Hiroshi Okada}
\email{hiroshi3okada@htu.edu.cn}
\affiliation{Department of Physics, Henan Normal University, Xinxiang 453007, China}

\date{\today}

\begin{abstract}
{
We search for predictability of lepton masses and mixing patterns of type-II seesaw scenario in a non-holomorphic modular $A_4$ symmetry recently proposed by Qu and Ding. 
We propose three types of minimum predictive models with different assignments of modular weight,
satisfying the neutrino oscillation data in Nufit 5.2.
The cosmological bound on the sum of neutrino mass is stringent to our models and CMB bound $\sum D_\nu\le0.12$ eV can be satisfied by one of three models playing an important role in discriminating them.
%
}
 %
 \end{abstract}
\maketitle
\newpage

\section{Introduction}

Clarifying the neutrino physics is one of the important tasks to understand our world which would be described beyond the standard model(SM). Especially, neutrino masses and mixing patterns in the lepton {sector} are totally different from the quark {one}; tiny neutrino masses and large mixing angles.
Therefore, these differences would underlie some unique mechanisms. In fact, 
there exist some promising mechanisms; canonical seesaw~\cite{Yanagida:1979gs, Minkowski:1977sc, Mohapatra:1979ia}, type-II seesaw~\cite{Magg:1980ut, Cheng:1980qt} inverse seesaw~\cite{Mohapatra:1986bd, Wyler:1982dd}, linear seesaw~\cite{Wyler:1982dd, Akhmedov:1995ip, Akhmedov:1995vm}, radiative seesaw~\cite{Zee:1980ai, Ma:2006km, Kajiyama:2013zla, Zee:1985id, Babu:1988ki, Krauss:2002px, Aoki:2008av, Gustafsson:2012vj}, and so on.
In addition to these mechanisms, flavor symmetries are often applied in order to predict neutrino mass and their mixing patterns.
In particular, a breakthrough idea has been proposed by Qu and Ding~\cite{Qu:2024rns},
in which they have shown that modular flavor symmetries can be applied in a non-supersymmetric framework. It suggests that {\it we would describe predictive and(or) reconstructive mass matrices and their mixing patterns of flavor physics in a more realistic and reliable theory}, even though original modular symmetries still be verifiable and strong predictions in models~\cite{Feruglio:2017spp, Criado:2018thu, Kobayashi:2018scp, Okada:2018yrn, Nomura:2019jxj, Okada:2019uoy, deAnda:2018ecu, Novichkov:2018yse, Nomura:2019yft, Okada:2019mjf,Ding:2019zxk, Nomura:2019lnr,Kobayashi:2019xvz,Asaka:2019vev,Zhang:2019ngf, Gui-JunDing:2019wap,Kobayashi:2019gtp,Nomura:2019xsb, Wang:2019xbo,Okada:2020dmb,Okada:2020rjb, Behera:2020lpd, Behera:2020sfe, Nomura:2020opk, Nomura:2020cog, Asaka:2020tmo, Okada:2020ukr, Nagao:2020snm, Okada:2020brs,Kang:2022psa, Ding:2024fsf, Ding:2023htn, Nomura:2023usj, Kobayashi:2023qzt, Petcov:2024vph, Kobayashi:2023zzc, Nomura:2024ghc}.
{Moreover we can reduce field contents considering non-supersymmetric framework. For example, we need to introduce two isospin triplet superfields in type-II seesaw model in order to cancel gauge anomaly.  It is worth analyzing neutrino models under non-holomorphic modular symmetry as we expect different prediction since some modular forms are different from holomorphic one.}

In our paper, we apply a type-II seesaw mechanism in addition to a non-holomorphic modular $A_4$ symmetry. We show three types of minimum models {which are distinguished from a neutrino mass model via Weinberg operator 
analyzed by Qu and Ding~\cite{Qu:2024rns}}, satisfying the neutrino oscillation data in Nufit 5.2~\cite{Esteban:2020cvm}. 
Then, we demonstrate each the predictions through chi-square numerical analysis.
Interestingly, when we impose {cosmological bound on the sum of neutrino mass}, an only one of three models with normal hierarchy can be survived from CMB bound.

This paper is organized as follows.
In Sec. \ref{sec:II}, 
we show our setup of the modular $A_4$ assignments in the framework of type-II seesaw. 
In Sec. \ref{sec:III}, we provide chi-square numerical analyses for three types of minimum models and show each of prediction.
Finally, we summarize and conclude in Sec. \ref{sec:IV}.

\section{Model setup}
\label{sec:II}

\begin{table}[t!]
\begin{tabular}{|c||c|c|c|c|}\hline\hline  
& ~$\overline{ L_L}$~ & ~$ \ell_R$~ & ~$ H$ ~&~ {$\Delta^\dagger$}~  \\\hline\hline 
$SU(2)_L$   & $\bm{2}$  & $\bm{1}$  & $\bm{2}$  & $\bm{3}$     \\\hline 
$U(1)_Y$    & $\frac12$  & $-1$ & $\frac12$ & {$-1$}    \\\hline
$A_4$   & $\bm{3}$  & $ \{ \bm{1} \} $  & $\bm{1}$ & $\bm{1}$         \\\hline 
$-k_I$    & $+1$  & $r$ & $0$ & $+2$      \\\hline
\end{tabular}
\caption{Charge assignments of the SM leptons $\overline{ L_L}$ and $\ell_R$ and $\Delta$
under $SU(2)_L\otimes U(1)_Y \otimes A_4$ where $-k_I$ is the number of modular weight. Here {$\{ \bm{1} \} =\{1, 1', 1''\}$} indicates assignment of $A_4$ singlets.
We consider three cases of $r=(-1,+1,+3)$.
}\label{tab:1}
\end{table}

Here, we review our model setup.
We assign the left-handed leptons $\overline{ L_L}$ to be triplet under $A_4$ with $+1$ modular weight.
We assign the right-handed leptons ${ \ell_R}$ to be three different types of singlets $(1,1',1'')$~\footnote{If we assign triplet instead of three types of triplets, we would not find any solutions to satisfy the current neutrino oscillations in Nufit 5.2~\cite{Esteban:2020cvm}.} under $A_4$ with $+r$ modular weight where $r$ is a positive integer.
In addition, we introduce  an isospin triplet scalar field $\Delta$ to generate neutrino mass matrix that is assigned by $A_4$ singlet with $+2$ modular weight . This field is requested {to realize type-II seesaw mechanism}. $H$ is denoted by the SM Higgs that is totally neutral under the modular symmetry.
Their fields and assignments are summarized in  Table~\ref{tab:1}.
Here, we search for some predictions in three {models characterized by} $r=(-1,+1,+3)$ that are minimum assignments to satisfy the neutrino oscillation data in Nufit 5.2~\cite{Esteban:2020cvm}.
The renormalizable Lagrangian under the modular $A_4$ is found as
\begin{align}
{\cal L}_\ell =
 [Y^{(-1-r)}_{3}\overline{ L_L} H  \ell_R] 
+
{[Y^{(-4)}_{1,3}\overline{ L_L}  \Delta^\dag (i\tau_2)  L^c_L]} +{\rm h.c.},
\label{eq:lpy}
\end{align}
where $\tau_2$ is the second Pauli matrix, $Y_{A}^{(k_Y)}$ shows a non-holomorphic modular form with $A_4$ representation $A$ and modular weight $k_Y$, and $[\cdots]$ represents a trivial $A_4$ singlet ${\bm 1}$ {constructed by fields and a modular form inside. 
We denote three models as Model04, Model24 and Model44 corresponding to choice of $r=-1$, $+1$ and $+3$ (two numbers associated with models are that of weights for modular forms in Yukawa Lagrangian).}

\subsection{Charged-lepton mass matrix}
The mass matrix of charged-lepton originates from the first term in Eq.(\ref{eq:lpy}).
After spontaneous electroweak symmetry breaking, the mass matrix is given by
\begin{align}
m_\ell = \frac{v_H}{\sqrt2}
 \left(\begin{array}{ccc} y_1^{(-1-r)} & y_3^{(-1-r)} & y_2^{(-1-r)} \\
 y_3^{(-1-r)} & y_2^{(-1-r)} & y_1^{(-1-r)} \\
  y_2^{(-1-r)} & y_1^{(-1-r)} & y_3^{(-1-r)} \end{array} \right)
   \left(\begin{array}{ccc} a_e & 0 & 0 \\
0 & b_e & 0 \\
0 & 0 & c_e \end{array} \right),
\label{eq:cgd-lep}
\end{align}
where $v_H$ is vacuum expectation value (VEV) of $H$, and $\{a_e, b_e, c_e\}$ are real parameters without loss of generality and used to fix the charged-lepton mass observables as can be seen below.
$Y^{(-1-r)}_{3}\equiv [y_1^{(-1-r)},y_2^{(-1-r)},y_3^{(-1-r)}]^T$~\footnote{In details, see ref.~\cite{Qu:2024rns}.}.
The mass eigenvalues for the charged-leptons are simply obtained via diag.$(m_e,m_\mu,m_\tau)\equiv V_L^\dag m_\ell V_R$.
Therefore, $V_L^\dag m_\ell m_\ell^\dag V_L ={\rm diag.}(|m_e|^2,|m_\mu|^2,|m_\tau|^2)$.
$\{a_e, b_e, c_e\}$ are determined in order to fit the three observed charged-lepton masses by the following relations:
\begin{align}
&{\rm Tr}[m_\ell m_\ell^\dag] = |m_e|^2 + |m_\mu|^2 + |m_\tau|^2,\quad
 {\rm Det}[m_\ell m_\ell^\dag] = |m_e|^2  |m_\mu|^2  |m_\tau|^2,\nn\\
&({\rm Tr}[m_\ell m_\ell^\dag)^2 -{\rm Tr}[(m_e m_e^\dag)^2] =2( |m_e|^2  |m_\mu|^2 + |m_\mu|^2  |m_\tau|^2+ |m_e|^2  |m_\tau|^2 ).\label{eq:l-cond}
\end{align}

\subsection{Neutrino mass matrix}
The mass matrix of neutrino comes from the second term in Eq.(\ref{eq:lpy}).
After spontaneous electroweak symmetry breaking, {$\Delta$ develops a VEV ($v_\Delta$), and} the mass matrix is given by
\begin{align}
m_\nu = \frac{a_\nu v_{\Delta}}{\sqrt2}
 \left(\begin{array}{ccc} 
 b_\nu + 2 y_1^{(-4)} & -y_3^{(-4)} & -y_2^{(-4)} \\
-y_3^{(-4)} & 2 y_2^{(-4)} & b_\nu - y_1^{(-4)}  \\
-y_2^{(-4)} &  b_\nu - y_1^{(-4)}&2 y_3^{(-4)}  \end{array} \right),
\label{eq:cgd-lep2}
\end{align}
where $\{a_\nu,b_\nu,$ are complex free parameters and $Y^{(-4)}_1$ is included by $a_\nu$.
Note here that $v_{\Delta}$ is required to be $v_\Delta \lesssim {\cal O}(1)$ GeV to satisfy the constraint on the $\rho$ parameter~\cite{ParticleDataGroup:2020ssz}.
The mass eigenvalues for the active neutrinos $D_\nu = \{D_{\nu_1}, D_{\nu_2}, D_{\nu_3} \}$ are obtained by $D_\nu \equiv V_\nu^\dag m_\nu V_\nu^*$.
Therefore, $V_\nu^\dag m_\nu m_\nu^\dag V_\nu ={\rm diag.}(|D_{\nu_1}|^2,|D_{\nu_2}|^2,|D_{\nu_3}|^2)$ where $D_{\nu_{1,2,3}}$ are the neutrino mass eigenvalues.
Pontecorvo-Maki-Nakagawa-Sakata (PMNS) mixing matrix $U(\equiv U_{PMNS})$ is defined by $U=V_L^\dag V_\nu$.
Here, we define dimensionless neutrino mass matrix; $m_\nu \equiv\kappa \tilde m_\nu$, $\kappa\equiv \frac{a_\nu v_{\Delta}}{\sqrt2}$ being a flavor independent  mass dimensional parameter. 
Therefore, we can rewrite $\kappa$ in terms of rescaled neutrino mass eigenvalues $\tilde D_\nu(\equiv D_\nu/\kappa)$ and atmospheric neutrino
mass-squared difference $\Delta m_{\rm atm}^2$ as follows:
\begin{align}
({\rm NH}):\  \kappa^2= \frac{|\Delta m_{\rm atm}^2|}{\tilde D_{\nu_3}^2-\tilde D_{\nu_1}^2},
\quad
({\rm IH}):\  \kappa^2= \frac{|\Delta m_{\rm atm}^2|}{\tilde D_{\nu_2}^2-\tilde D_{\nu_3}^2},
 \end{align}
where 
NH and IH stand for {the} normal
and inverted hierarchies, respectively. Subsequently, the solar neutrino mass-squared difference is described by
\begin{align}
\Delta m_{\rm sol}^2= {\kappa^2}({\tilde D_{\nu_2}^2-\tilde D_{\nu_1}^2}).
 \end{align}
This should be within the range of the experimental value. Later, we will adopt NuFit 5.2~\cite{Esteban:2020cvm} to our numerical analysis. 
The {effective mass for} neutrinoless double beta decay is given by 
\begin{align}
\langle m_{ee}\rangle=\kappa|\tilde D_{\nu_1} \cos^2\theta_{12} \cos^2\theta_{13}+\tilde D_{\nu_2} \sin^2\theta_{12} \cos^2\theta_{13}e^{i\alpha_{2}}+\tilde D_{\nu_3} \sin^2\theta_{13}e^{i(\alpha_{3}-2\delta_{CP})}|.
\end{align}
%

 Its predicted value is constrained by the current KamLAND-Zen data and could be measured in future~\cite{KamLAND-Zen:2024eml}.
The upper bound is found as $\langle m_{ee}\rangle<(36-156)$ meV at 90 \% confidence level where the range of the bound comes from the use of different method estimating nuclear matrix elements. 
Sum of neutrino masses is constrained by the minimal cosmological model
$\Lambda$CDM $+\sum D_{\nu_i}$ that provides the upper bound on $\sum D_{\nu}\le$ 120 meV~\cite{Vagnozzi:2017ovm, Planck:2018vyg}, although it becomes weaker if the data are analyzed in the context of extended cosmological models~\cite{ParticleDataGroup:2014cgo}.
Recently, DESI and CMB data combination provides more stringent upper bound on the sum of neutrino masses~$\sum D_{\nu}\le$ 72 meV~\cite{DESI:2024mwx}. 
%
The two observable $\langle m_{ee}\rangle$ and $\sum D_{\nu_i}$ are also taken into account in the numerical analysis.

\section{Numerical analysis and phenomenology}
\label{sec:III}
In this section, we carry out numerical analysis to fit the neutrino data and find some predictions of the models.
Then some implications for phenomenology are discussed.
We randomly select our input parameters within the following ranges 
\begin{align}
 \tau_{\rm Re}\in [-0.5,0.5], \quad \tau_{\rm Im}\in \left[\frac{\sqrt3}{2}, 5\right], \quad |b_\nu| \in [10^{-3},10^{3}], \quad {\rm arg}[b_\nu] \in [-\pi,\pi].
\end{align}

\begin{figure}[tb]
\begin{center}
\includegraphics[width=50.0mm]{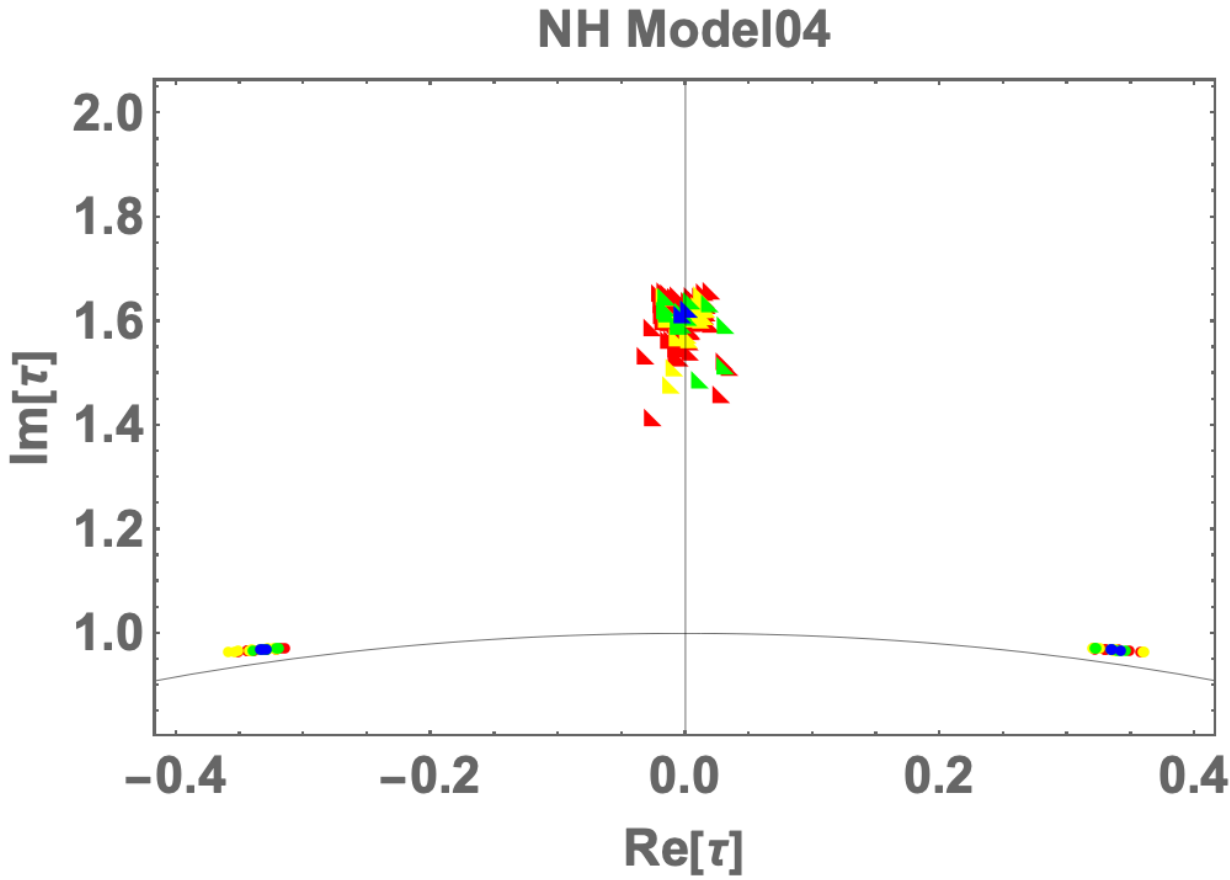} \quad
\includegraphics[width=50.0mm]{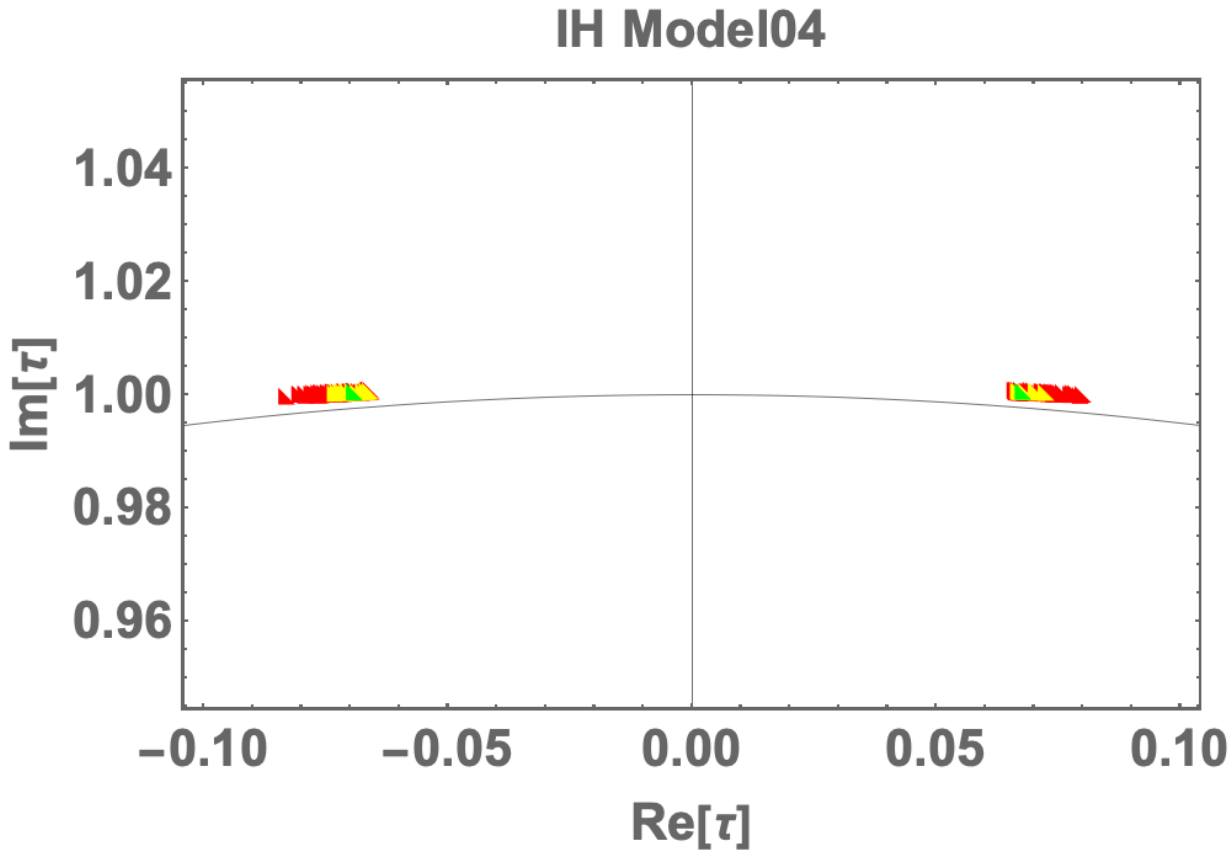} \quad
\caption{Numerical $\Delta\chi^2$ analyses in case of $r=-1$, where the blue color represents the range of $0-1$, the green $1-2$, the yellow $2-3$, and the red $3-5$ {$\sigma$ standard deviation estimated by} $\sqrt{\Delta\chi^2}$. The black solid line is the boundary of the fundamental domain at $|\tau|=1$. The left figure represents the case of NH and the right one IH. Circle points satisfy $\sum D_\nu\le$ 120 meV, otherwise triangle plots are depicted.  }
  \label{fig:tau04}
\end{center}\end{figure}
%
\subsection{$r=-1$}
In Fig.~\ref{fig:tau04}, we figure out the allowed range of $\tau$ in case of $r=-1$
where the left figure represents the case of NH and the right one IH.
The blue color represents the range of $0-1$, the green $1-2$, the yellow $2-3$, and the red $3-5$ {$\sigma$ standard deviation estimated by} $\sqrt{\Delta\chi^2}$. The black solid line is the boundary of the fundamental domain at $|\tau|=1$. Circle points satisfy $\sum D_\nu\le$ 120 meV, otherwise triangle plots are depicted.
NH solutions are localized at nearby two regions $\tau=1.5 i$ and $\tau=\pm0.35+0.96 i$,
while IH solutions are localized at nearby $\tau=\pm0.07+i$.

\begin{figure}[tb]
\begin{center}
\includegraphics[width=50.0mm]{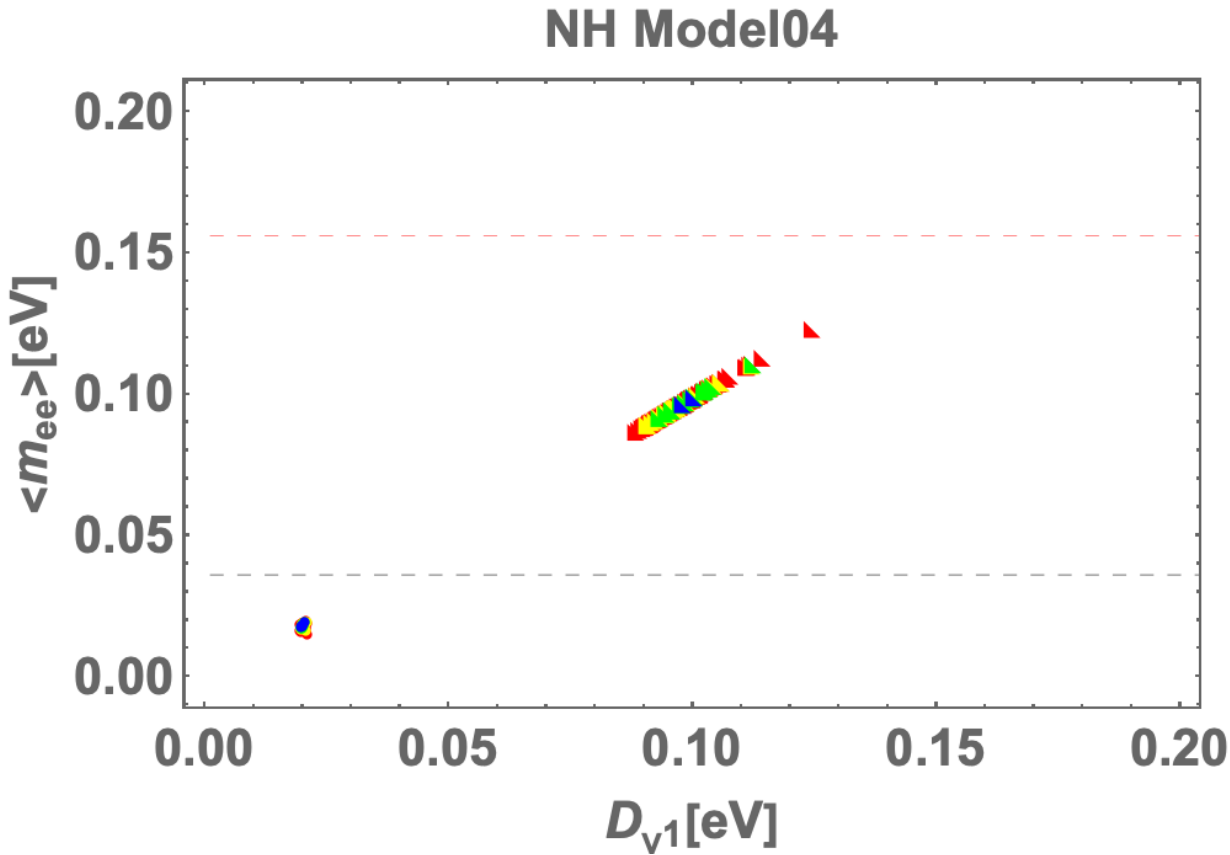} \quad
\includegraphics[width=50.0mm]{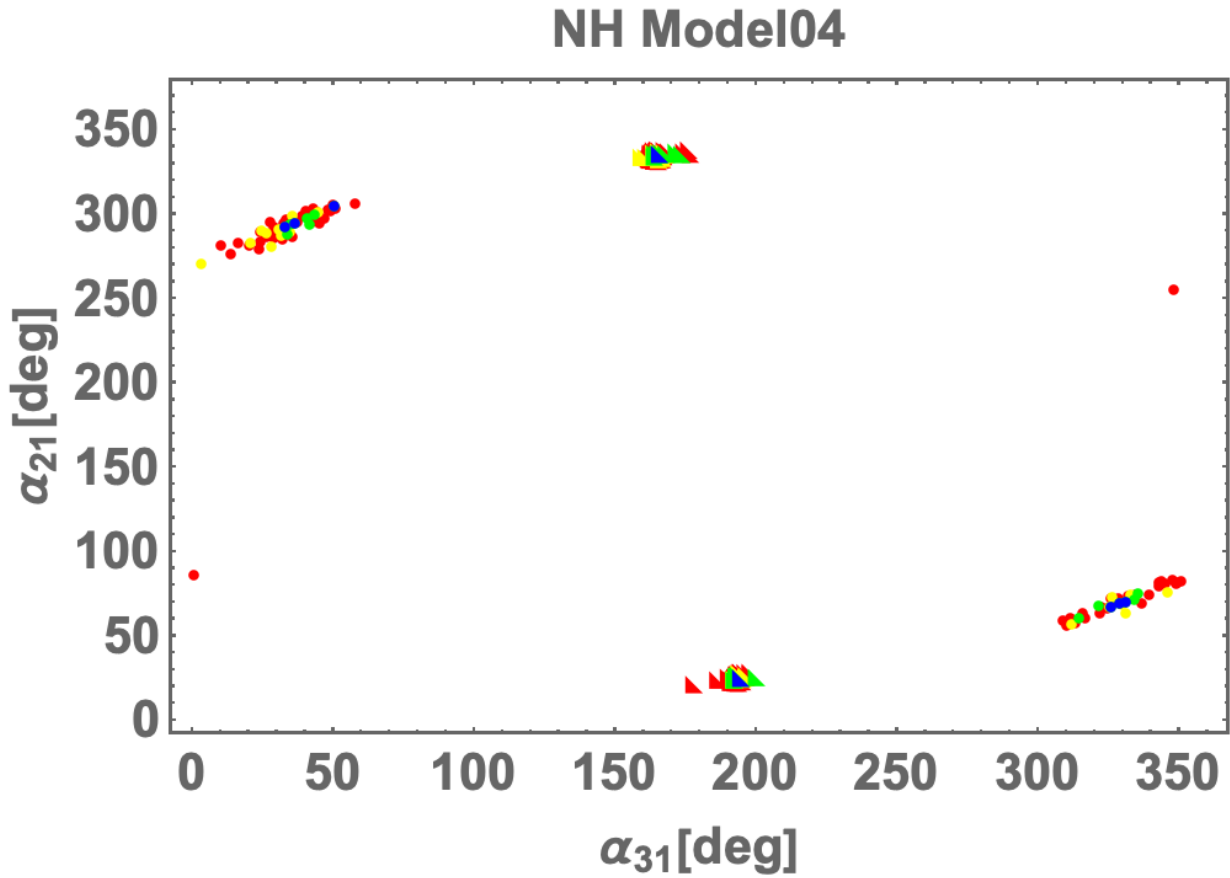}  \quad
\includegraphics[width=50.0mm]{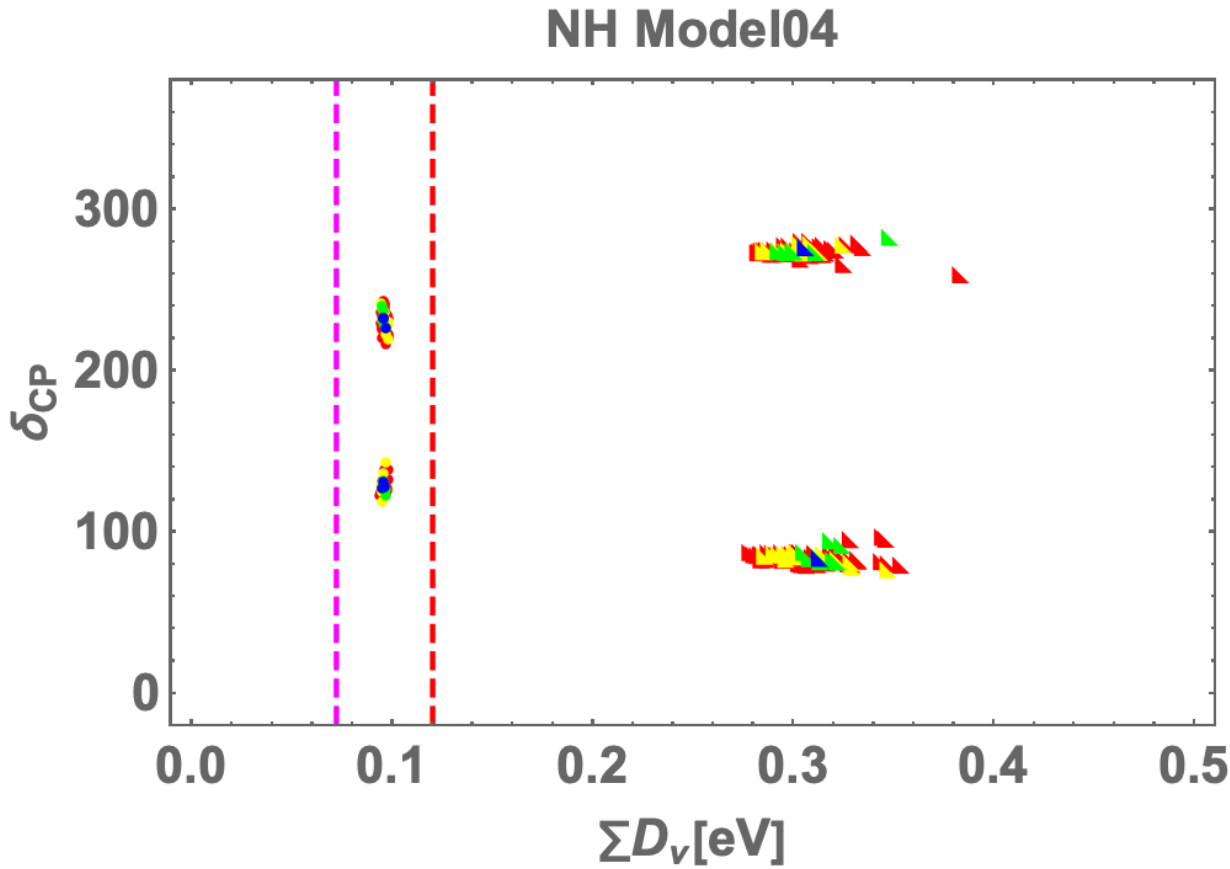} \\
\includegraphics[width=50.0mm]{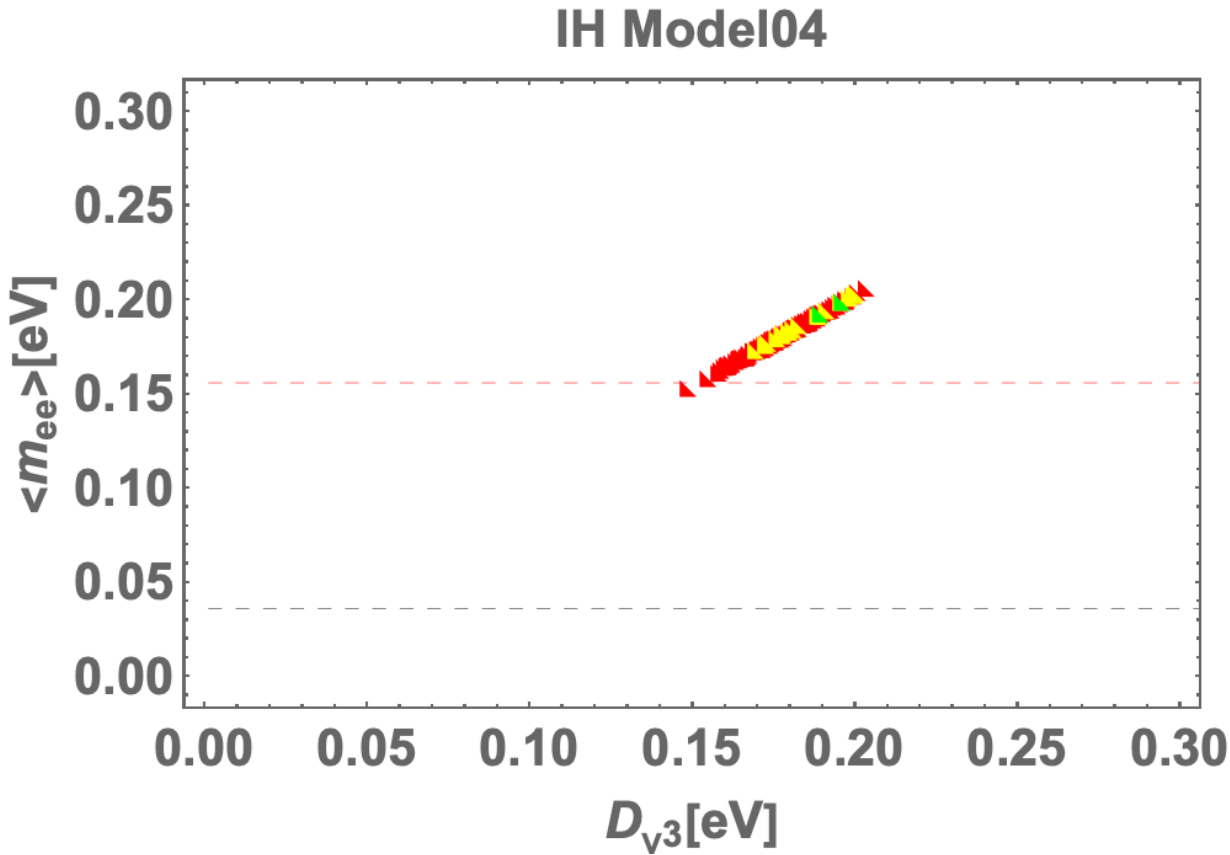} \quad
\includegraphics[width=50.0mm]{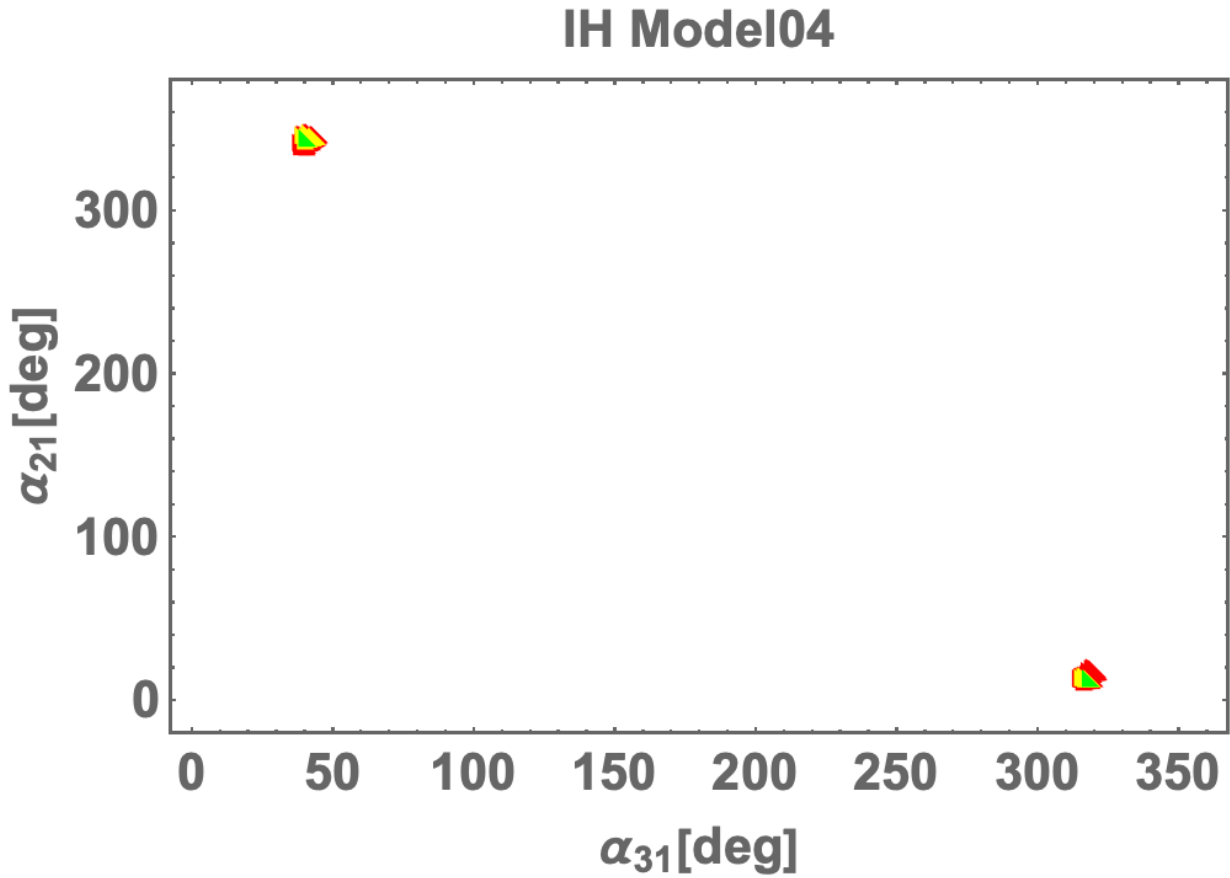}  \quad
\includegraphics[width=50.0mm]{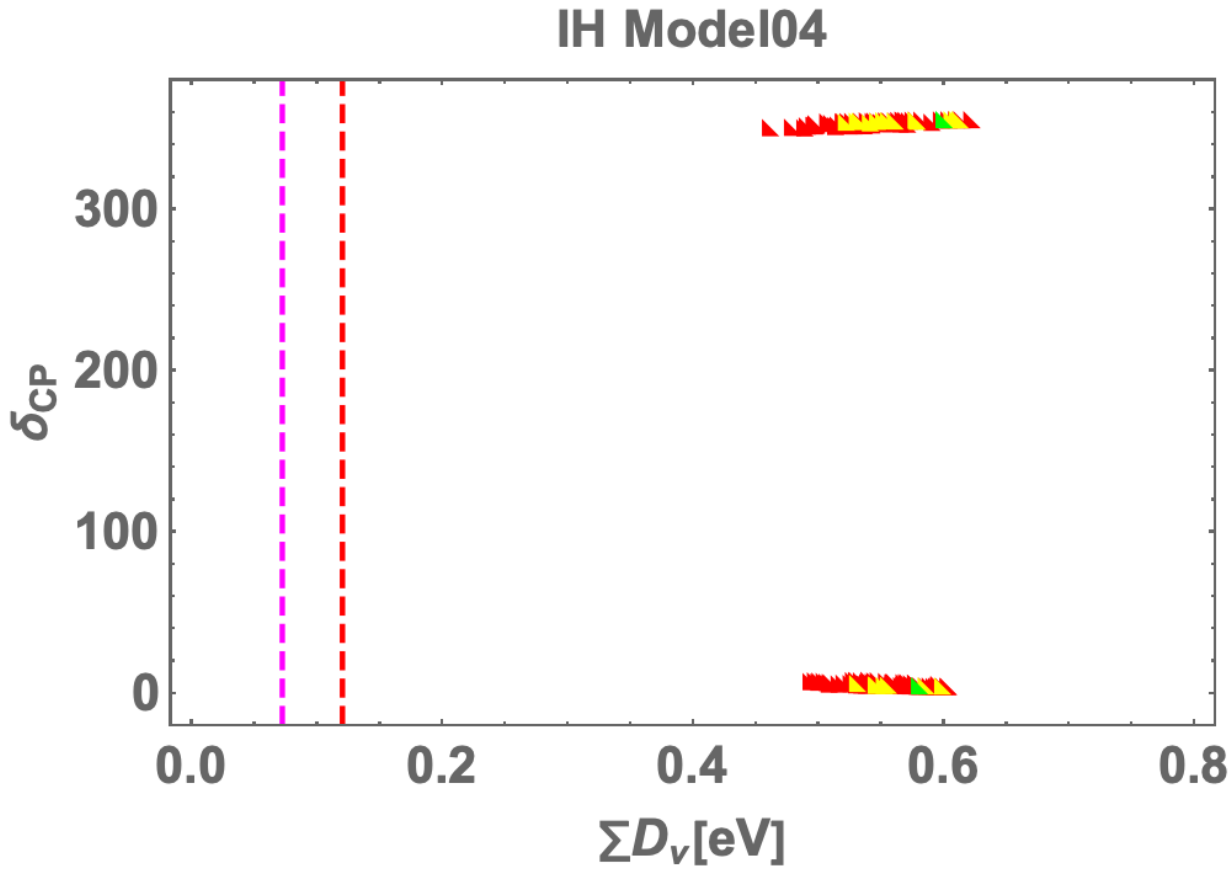} \\
\caption{Numerical analyses in case of $r=-1$.
The up and down figures represent the NH and the IH cases, respectively. 
The left one demonstrates neutrinoless double beta decay $\langle m_{ee}\rangle$ in terms of the lightest neutrino mass {where red dashed horizontal lines indicate the current upper bound from KamLAND-Zen with different estimation for nuclear matrix elements.}.
The center one shows Majorana phases $\alpha_{21}$ and $\alpha_{31}$.
The right one depicts the Dirac CP phase in terms of the sum of neutrino masses {where the red(magenta) dashed vertical lines indicate the upper bound on $\sum D_{\nu}$ from Planck CMB (+ DESI BAO) data}. The colors of plots and shapes are the same as the ones of Fig.~\ref{fig:tau04}.}
  \label{fig:04s}
\end{center}\end{figure}
%
In Fig.~\ref{fig:04s}, we show predictions in the case of $r=-1$.
The up and down figures represent the NH and the IH cases, respectively.
The plot-legends of color and shape  are the same as the ones of Fig.~\ref{fig:tau04}.
The left one demonstrates neutrinoless double beta decay $\langle m_{ee}\rangle$ in terms of the lightest neutrino mass.
The dotted horizontal lines are bounds from the current KamLAND-Zen $\langle m_{ee}\rangle<(36-156)$ meV as discussed in the neutrino sector.
In case of NH, there exist two localized regions; $D_{\nu_1}\approx(0.02,\ 0.09-0.13)$ eV and $\langle m_{ee}\rangle \approx(0.02,\ 0.08-0.12)$ eV.
In case of IH, we have allowed region at $D_{\nu_3}\approx(0.15-0.21)$ eV and $\langle m_{ee}\rangle \approx(0.15-0.21)$ eV.
Note here that the only NH satisfies {the CMB bound on the sum of neutrino mass.}
The center one shows Majorana phases $\alpha_{31}$ and $\alpha_{21}$.
There exist four main localized regions in case of NH;
$\alpha_{31}\approx(0-60,\ 160-200,\ 310-360)$ deg and $\alpha_{21}\approx(10-30,\ 50-80,\ 270-310,\ 330-350)$ deg.
There exist two localized regions in case of IH;
$\alpha_{31}\approx(35-50,\ 320-330)$ deg and $\alpha_{21}\approx(0-10,\ 330-340)$ deg.
The right one depicts the Dirac CP phase  in terms of the sum of neutrino masses .
The vertical red dashed line shows the upper bound of cosmological limit $\sum D_\nu\le0.12$ eV while the vertical magenta dashed line shows the upper bound of $\sum D_\nu\le0.072$ eV.
In case of NH, there exist four localized regions; sum $D_{\nu}\approx(0.1,\ 0.28-0.4)$ eV
and $\delta_{CP} \approx(80-100,\ 120-140,\ 220-240,\ 260-280)$ deg.
In case of IH, there exist two localized regions; sum $D_{\nu}\approx(0.5-0.65)$ eV
and $\delta_{CP} \approx(0-20,\ 340-360)$ deg.

\subsection{$r=+1$}
%
\begin{figure}[tb]
\begin{center}
\includegraphics[width=50.0mm]{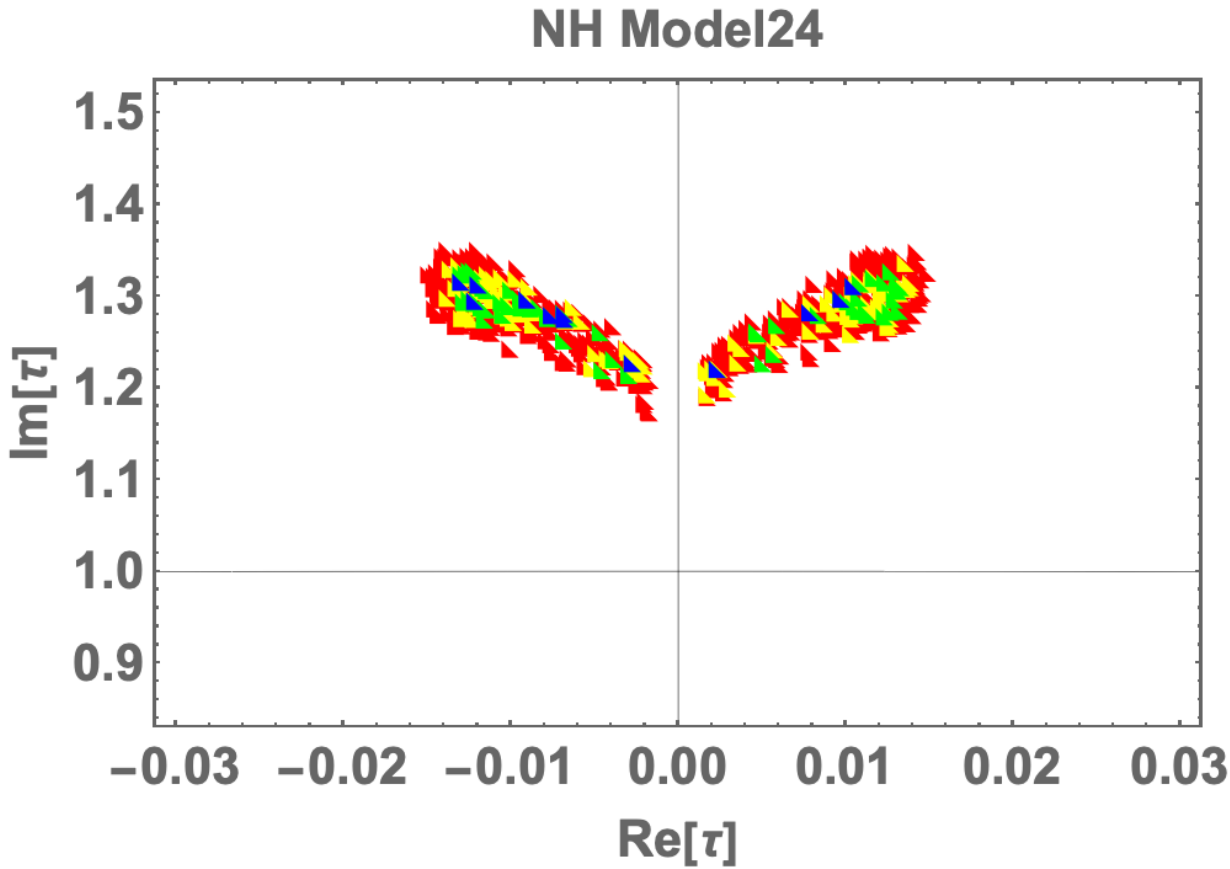} \quad
\caption{Numerical $\Delta\chi^2$ analyses in case of $r=+1$, where all the legends are the same as Fig.~\ref{fig:tau04}. We would not find any solutions in case of IH}
  \label{fig:tau24}
\end{center}\end{figure}
In Fig.~\ref{fig:tau24}, we figure out the allowed range of $\tau$ in case of $r=+1$
where we would not find any solutions in case of IH.
All the legends are the same as Fig.~\ref{fig:tau04}.
We have a specific region of $\tau$; $|$Re[$\tau$]$|$$\approx[0-0.016]$ and  
Im[$\tau$]$\approx[1.18-1.35]$.

\begin{figure}[tb]
\begin{center}
\includegraphics[width=50.0mm]{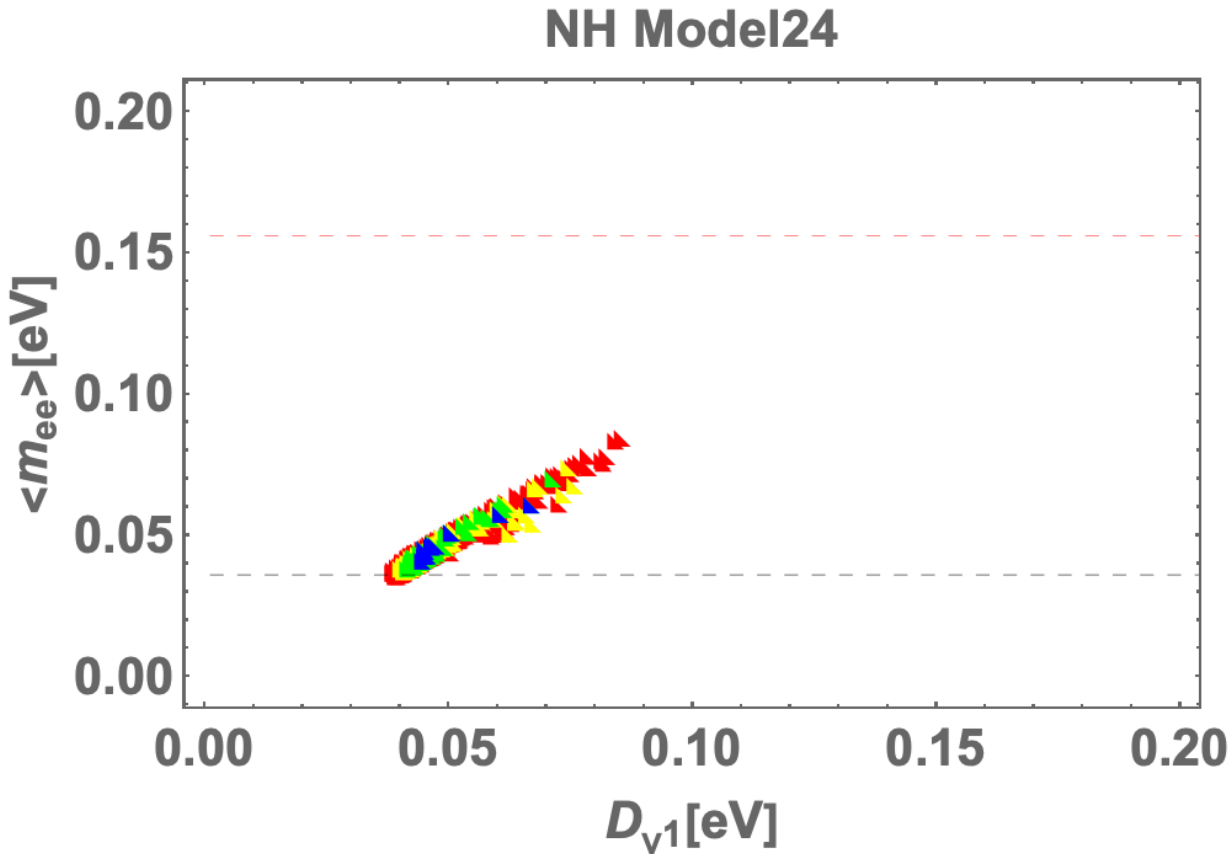} \quad
\includegraphics[width=50.0mm]{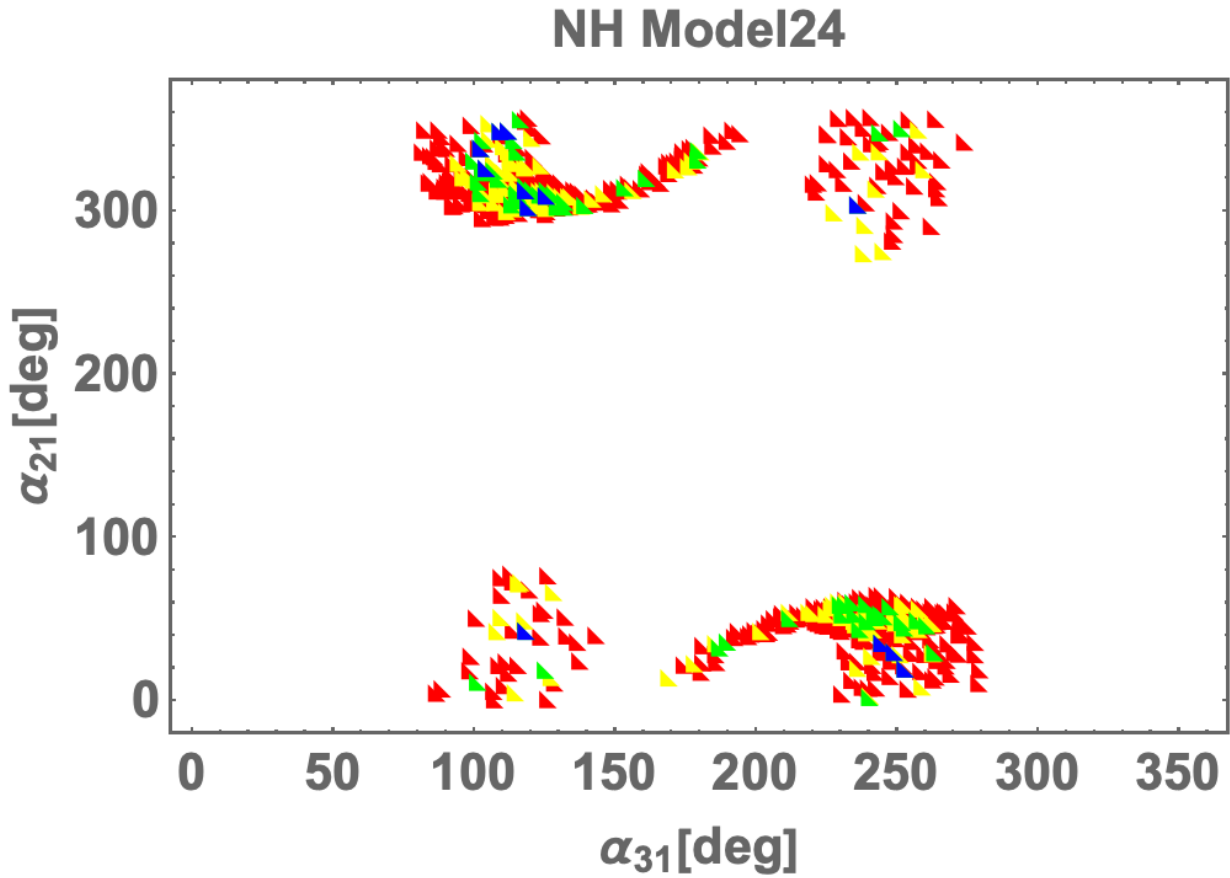}  \quad
\includegraphics[width=50.0mm]{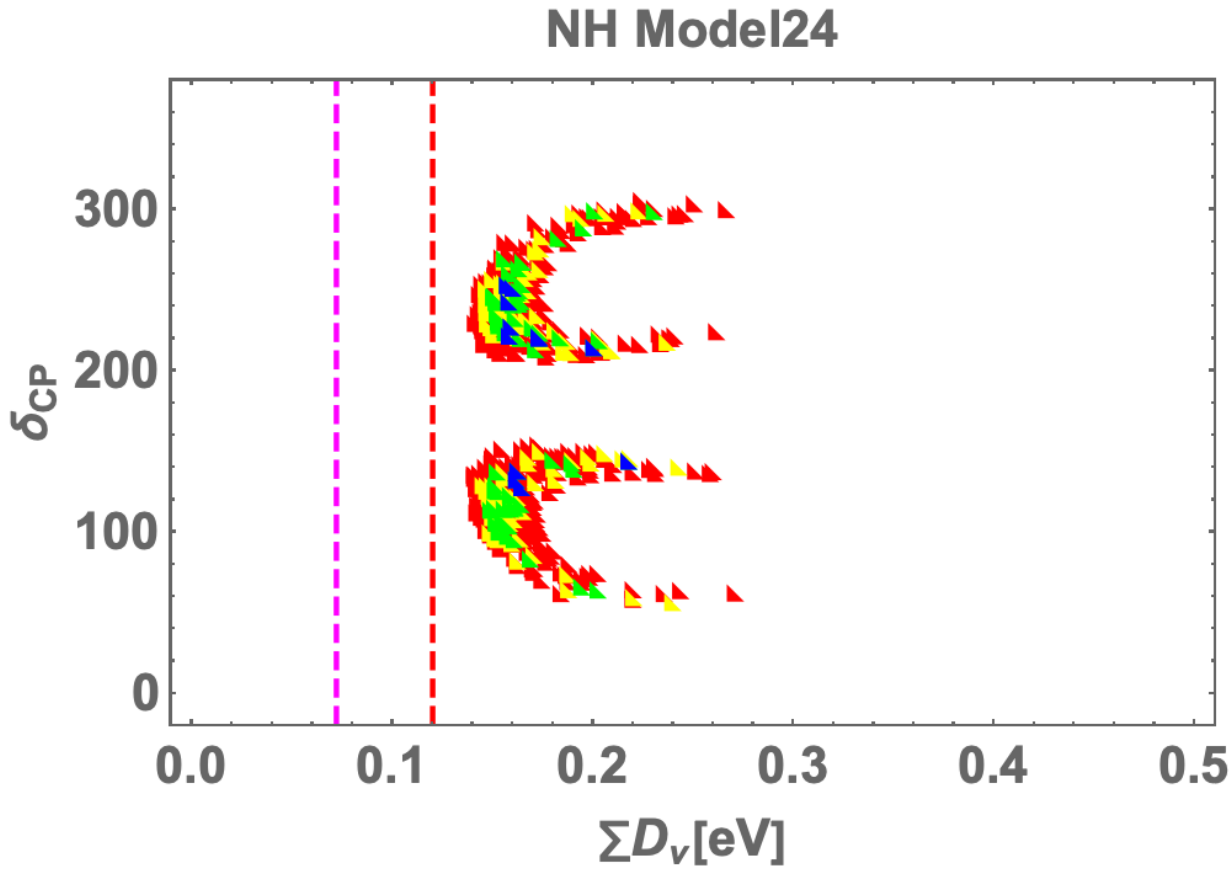} \\
\caption{Numerical $\Delta\chi^2$ analyses in case of $r=+1$, where
All the legends are the same as Fig.~\ref{fig:04s}.  We would not find any solutions in case of IH}
  \label{fig:24s}
\end{center}\end{figure}
%
In Fig.~\ref{fig:24s}, we show predictions in the case of $r=+1$, where
All the legends are the same as Figs.~\ref{fig:04s}.
We would not find any solutions in case of IH
The left figure suggests $D_{\nu_1}\approx(0.04-0.08)$ eV and $\langle m_{ee}\rangle \approx(0.04-0.08)$ eV.
The center one shows there exist four regions;
$\alpha_{31}\approx(80-280)$ deg and $\alpha_{21}\approx(0-80,\ 280-360)$ deg.
The right one shows that there exist two localized regions; sum $D_{\nu}\approx(0.14-0.28)$ eV
and $\delta_{CP}\rangle \approx(40-160,\ 200-300)$ deg.

\subsection{$r=+3$}
%
\begin{figure}[tb]
\begin{center}
\includegraphics[width=50.0mm]{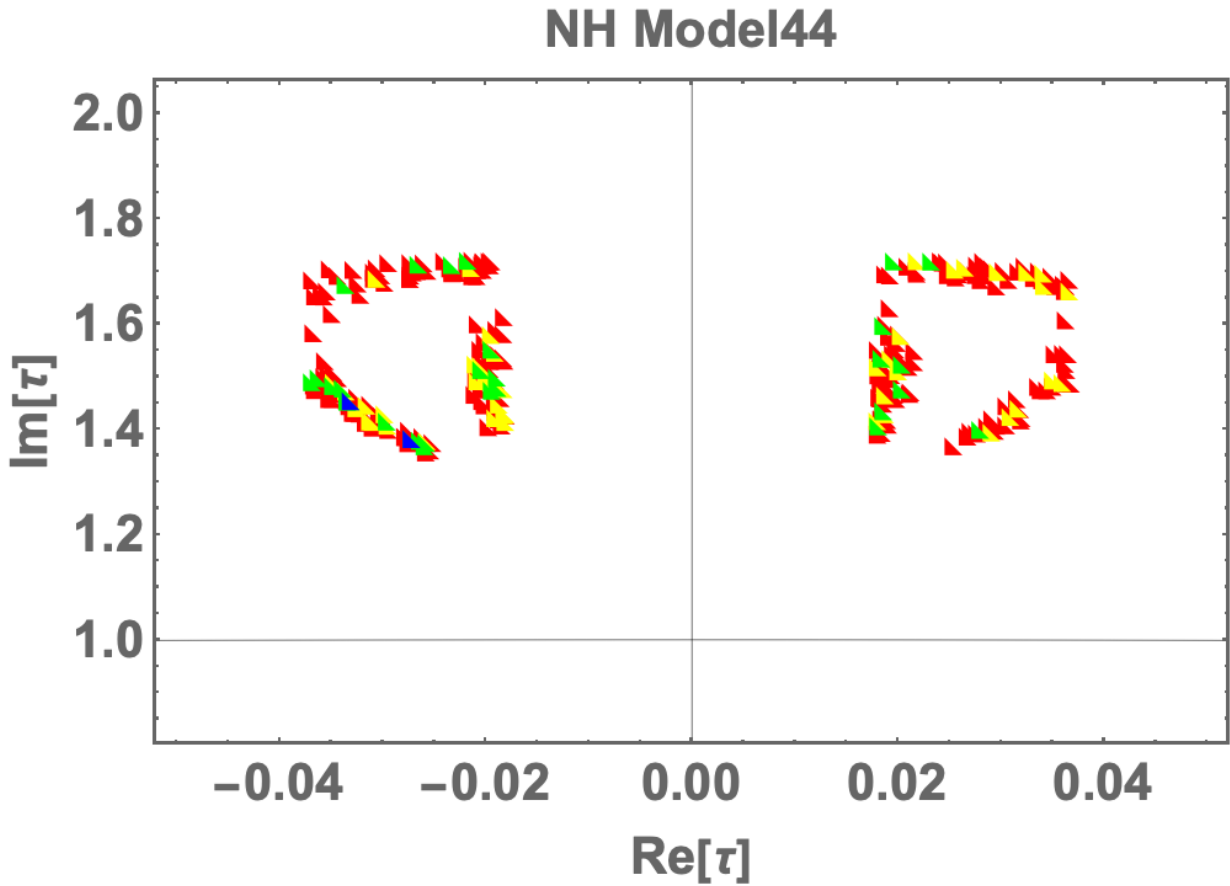} \quad
\includegraphics[width=50.0mm]{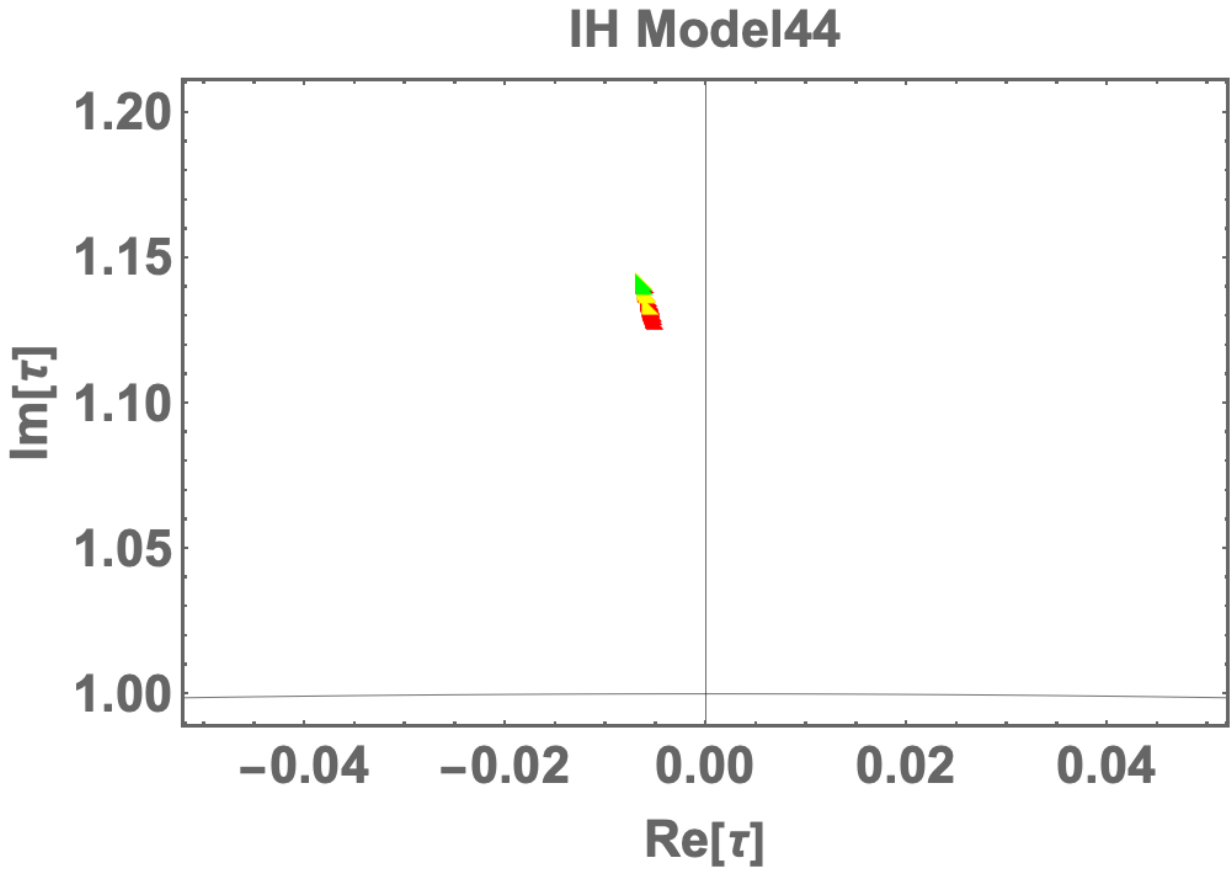} \quad
\caption{Numerical $\Delta\chi^2$ analyses in case of $r=+3$, where all the legends are the same as Fig.~\ref{fig:tau04}.}
  \label{fig:tau44}
\end{center}\end{figure}
In Fig.~\ref{fig:tau44}, we figure out the allowed range of $\tau$ in case of $r=+3$.
All the legends are the same as Fig.~\ref{fig:tau04}.
In case of NH, we have a unique allowed shape of $\tau$; $|$Re[$\tau$]$|$$\approx[0.018-0.04]$ and Im[$\tau$]$\approx[1.3-1.7]$.
In case of IH, we have a localized region of $\tau$; Re[$\tau$]$\approx0.002$ and Im[$\tau$]$\approx1.13$.

\begin{figure}[tb]
\begin{center}
\includegraphics[width=50.0mm]{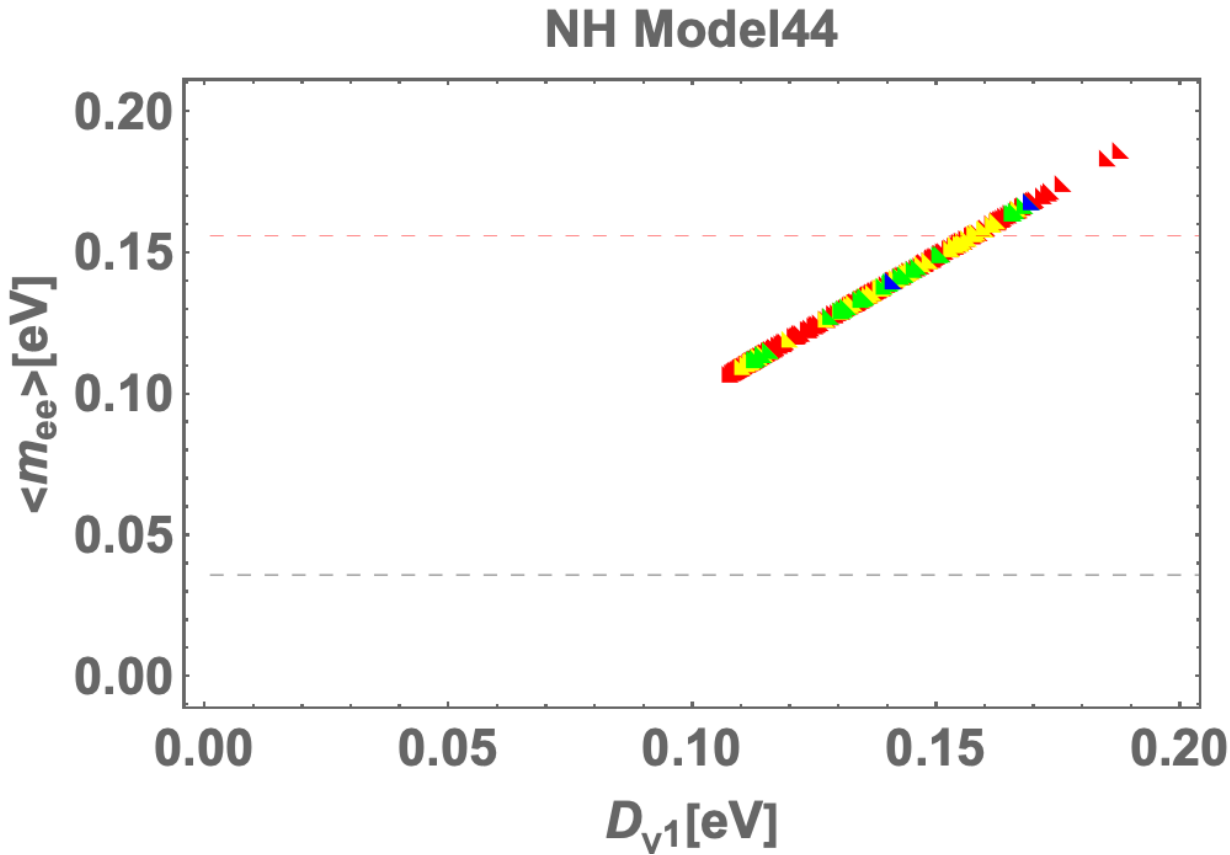} \quad
\includegraphics[width=50.0mm]{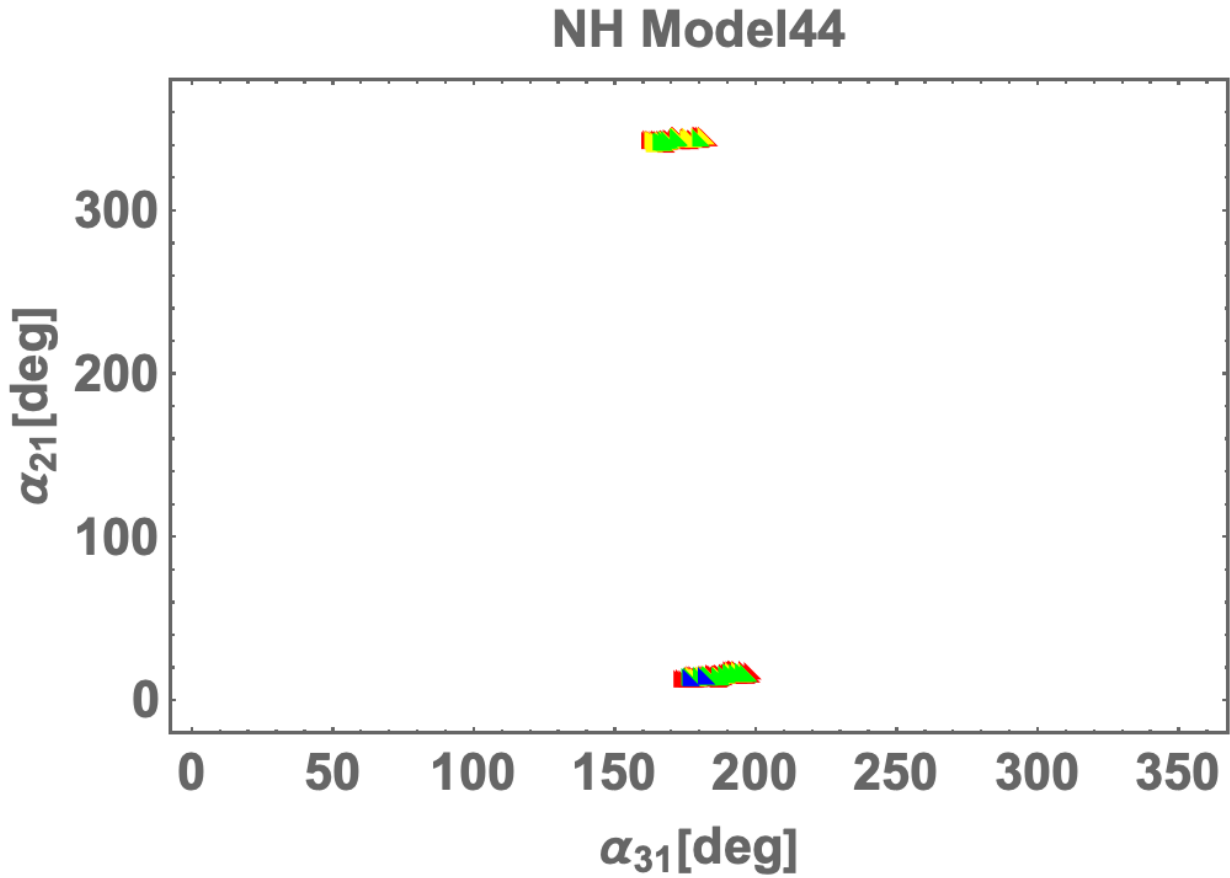}  \quad
\includegraphics[width=50.0mm]{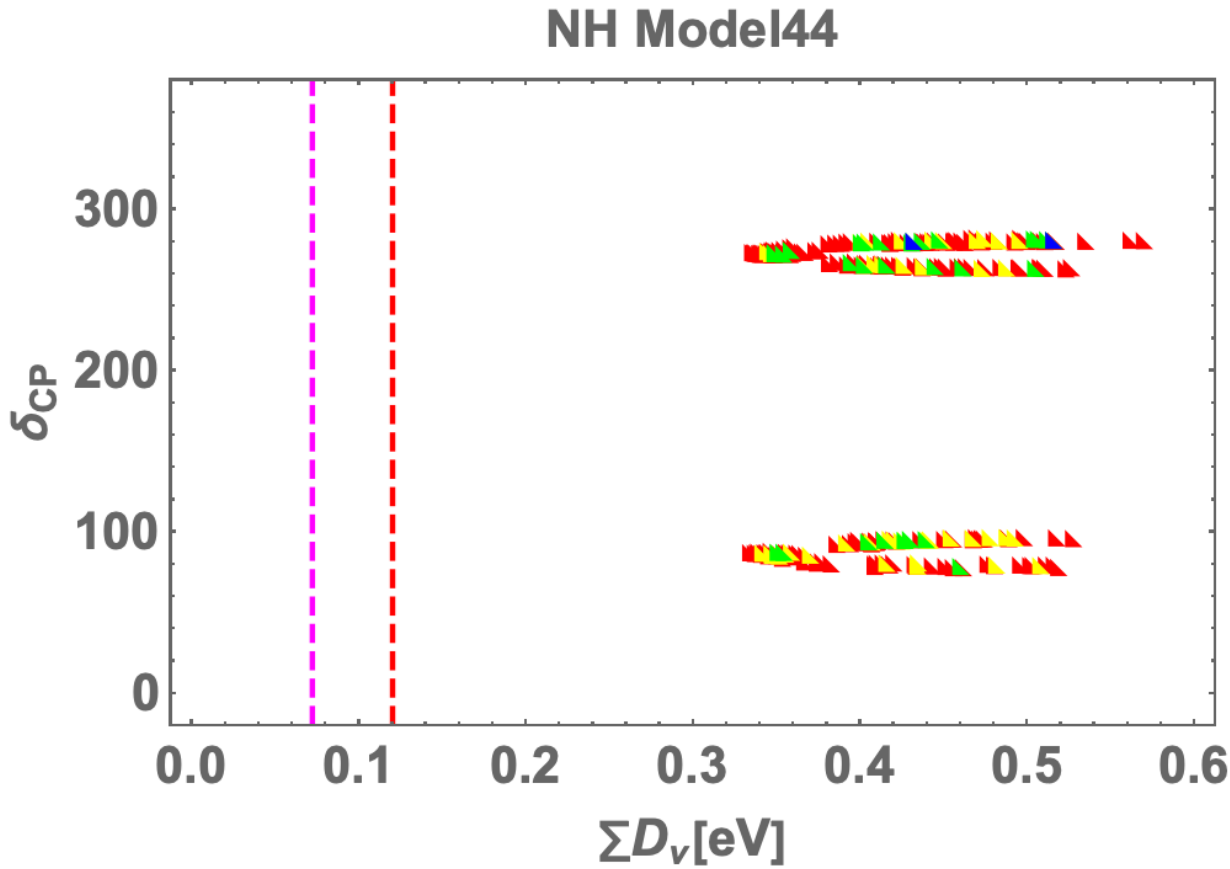} \\
\includegraphics[width=50.0mm]{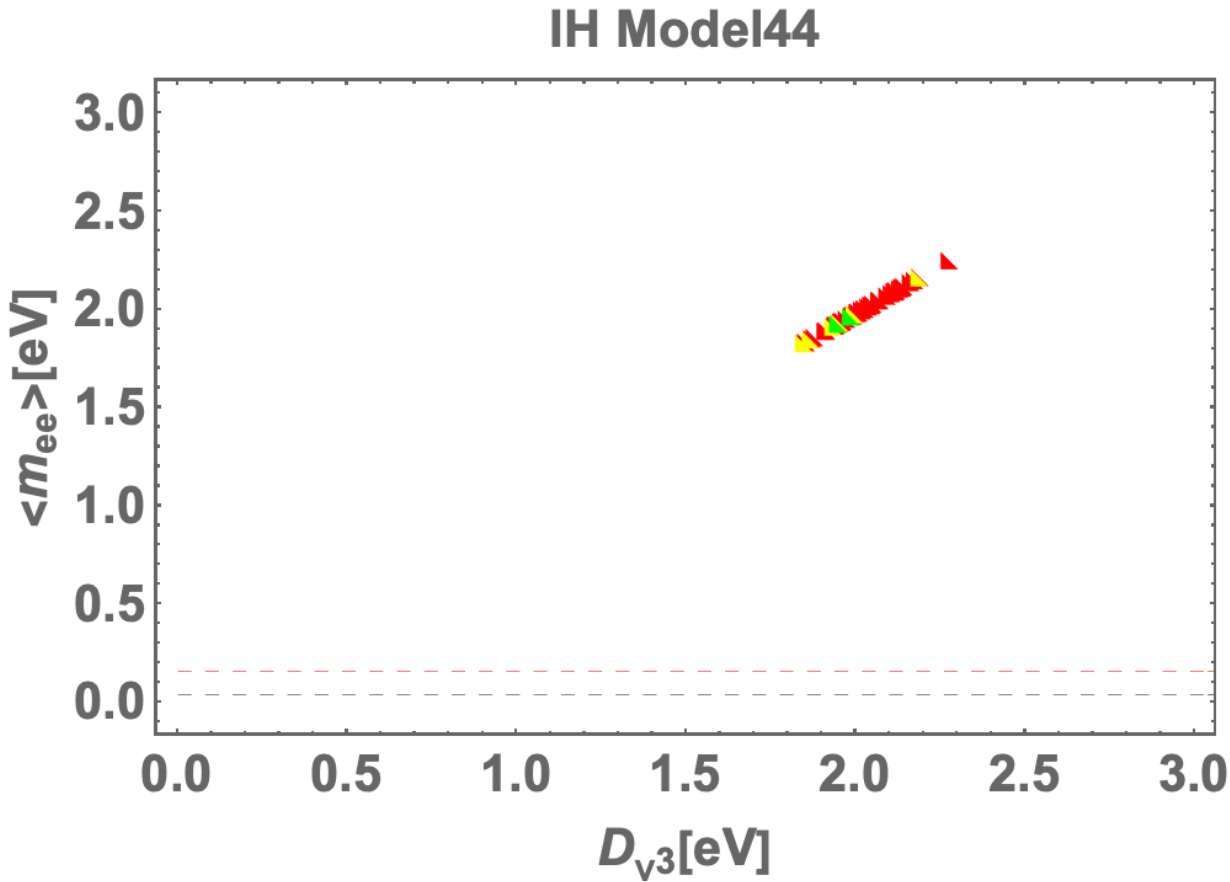} \quad
\includegraphics[width=50.0mm]{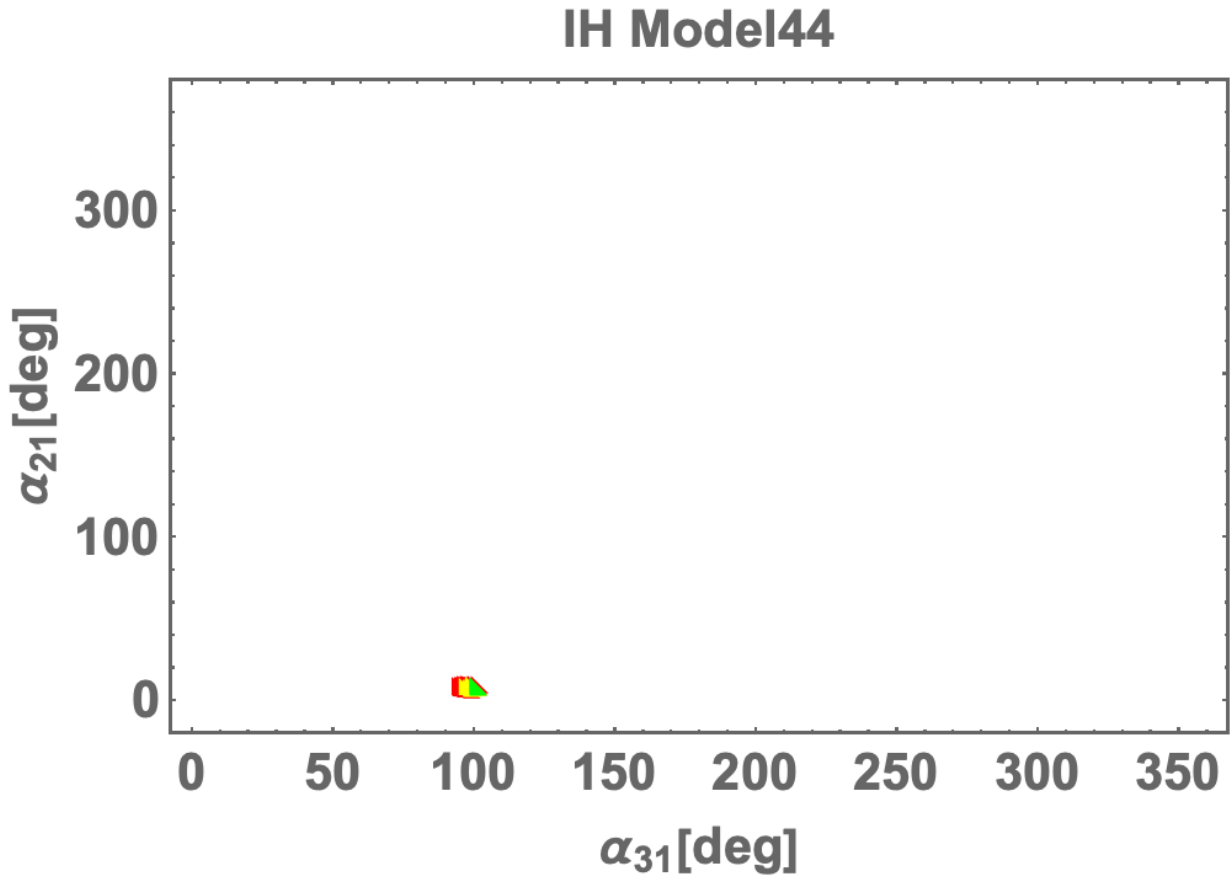}  \quad
\includegraphics[width=50.0mm]{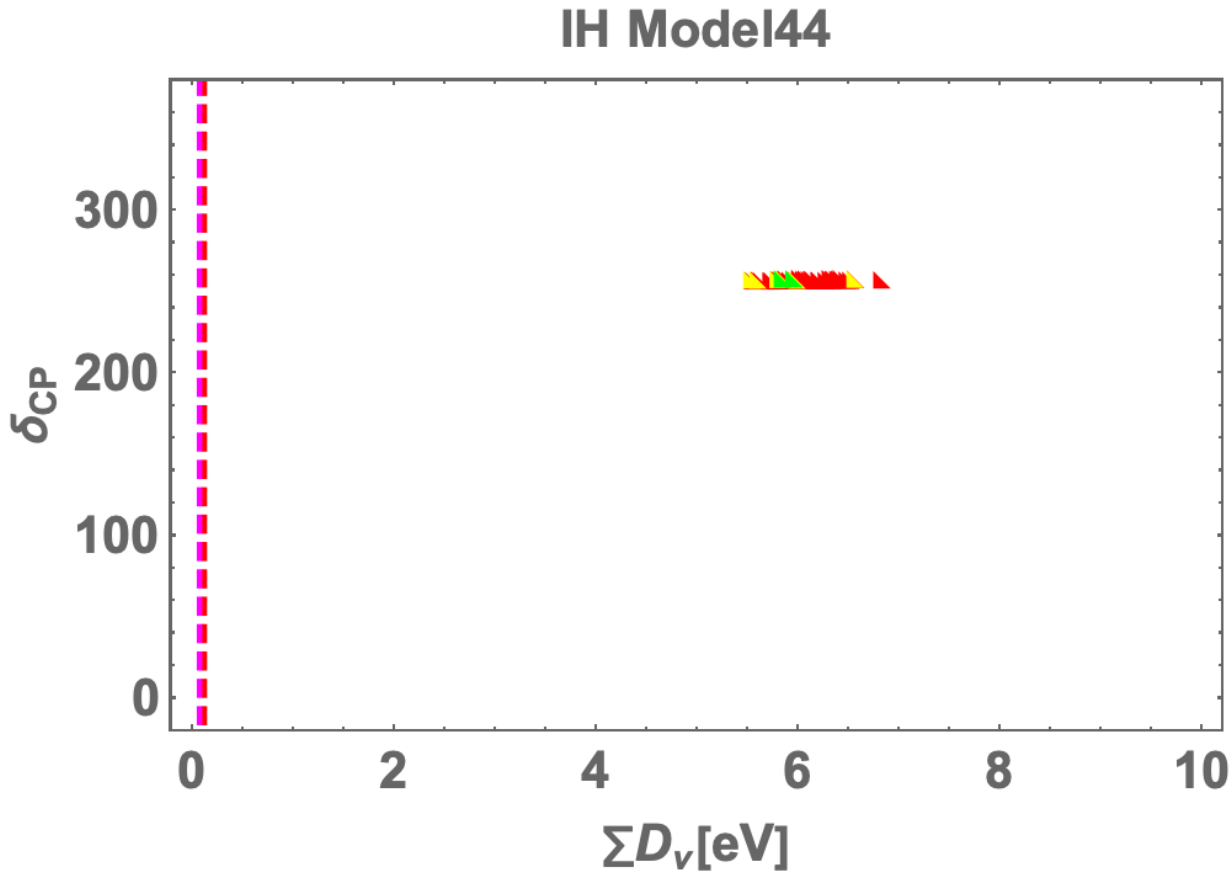} \\
\caption{Numerical $\Delta\chi^2$ analyses in case of $r=+1$, where
All the legends are the same as Fig.~\ref{fig:04s}. }
  \label{fig:44s}
\end{center}\end{figure}
%
In Fig.~\ref{fig:44s}, we show predictions in the case of $r=+3$, where
all the legends are the same as Fig.~\ref{fig:04s}.
%
In case of NH, we have the allowed regions at $D_{\nu_1}\approx(0.11-0.19)$ eV and $\langle m_{ee}\rangle \approx(0.1-0.18)$ eV.
In case of IH, we have allowed region at $D_{\nu_3}\approx(1.8-2.2)$ eV and $\langle m_{ee}\rangle \approx(1.8-2.2)$ eV.
There exist two localized regions in case of NH;
$\alpha_{31}\approx(150-200)$ deg and $\alpha_{21}\approx(0-10,\ 350-360)$ deg.
There exist a localized region in case of IH;
$\alpha_{31}\approx 100$ deg and $\alpha_{21}\approx 0$ deg.
%
In case of NH, there exist two localized regions; sum $D_{\nu}\approx(0.35-0.58)$ eV
and $\delta_{CP}\rangle \approx(80-100,\ 260-280)$ deg.
In case of IH, there exist a localized region; sum $D_{\nu}\approx(5.5-6.5)$ eV
and $\delta_{CP}\rangle \approx270$ deg.

\section{Summary and discussion}
\label{sec:IV}
We have searched for predictability of lepton masses and mixing of type-II scenario in a non-holomorphic modular $A_4$ symmetry. 
We have found three types of minimum models in addition to the Qu and Ding model,
satisfying the neutrino oscillation data in Nufit 5.2.
However, if we impose the cosmological bound $\sum D_\nu\le0.12$ eV, the only normal hierarchical model with $r=-1$ is survived. Furthermore, we have sharp predictions in this case as follows:
There are two localized regions; $D_{\nu_1}\approx(0.02,\ 0.09-0.13)$ eV and $\langle m_{ee}\rangle \approx(0.02,\ 0.08-0.12)$ eV.
We also found four main localized regions
$\alpha_{31}\approx(0-60,\ 160-200,\ 310-360)$ deg and $\alpha_{21}\approx(10-30,\ 50-80,\ 270-310,\ 330-350)$ deg.
There exist four localized regions; sum $D_{\nu}\approx(0.1,\ 0.28-0.4)$ eV
and $\delta_{CP}\rangle \approx(80-100,\ 120-140,\ 220-240,\ 260-280)$ deg.
This model is well-tested by upcoming experiments in near future.

\section*{Acknowledgments}
The work was supported by the Fundamental Research Funds for the Central Universities (T.~N.).

\bibliography{ctma4.bib}
\end{document}